# Conditional expectations associated with quantum states

Gerd Niestegge
*Zillertalstrasse 39, D-81373 Muenchen, Germany*

*Abstract.* An extension of the conditional expectations (those under a given sub-algebra of events and not the simple ones under a single event) from the classical to the quantum case is presented. In the classical case, the conditional expectations always exist; in the quantum case, however, they exist only if a certain weak compatibility criterion is satisfied. This compatibility criterion was introduced among others in a recent paper by the author. Then, state-independent conditional expectations and quantum Markov processes are studied. A classical Markov process is a probability measure, together with a system of random variables, satisfying the Markov property and can equivalently be described by a system of Markovian kernels (often forming a semi-group). This equivalence is partly extended to quantum probabilities. It is shown that a dynamical (semi-) group can be derived from a given system of quantum observables satisfying the Markov property, and the group generators are studied. The results are presented in the framework of Jordan operator algebras, and a very general type of observables (including the usual real-valued observables or self-adjoint operators) is considered.

## I. INTRODUCTION

A classical Markov process is a probability measure together with a system of random variables, satisfying the Markov property (which characterizes a certain memorilessness of the process). A classical Markov process can equivalently be described by a system of Markovian kernels forming a semi-group when the process is stationary.

The quantum analogues of probability measures, random variables and Markovian kernels are the states, observables and unital positive linear maps. A quantum Markov process is usually modeled as a semi-group of unital positive linear maps on a W*-algebra (von Neumann algebra). The major result of the present paper is the derivation of such a semi-group from a stochastic process given as a quantum state together with a system of observables, satisfying a certain directed weak compatibility criterion, the Markov property and the stationarity criterion. This partly extends the classical equivalence between stationary Markov processes and semi-groups of Markovian kernels to the quantum case. As in the classical case, the major tools are the Radon-Nikodym theorem and conditional expectations, the extension of which to the quantum case is a major result of the present paper.

The quantum probabilities are considered in the general framework of Jordan operator algebras, particularly the so-called JBW algebras. A justification for this is given in Ref. 10. An observable then becomes a homomorphism between such algebras. A classical random variable can be considered an observable on a classical space with values in a classical space. A real-valued quantum observable (spectral measure of a self-adjoint operator) can be considered an observable on a non-classical quantum space with values in a classical space (the real numbers). This paper, however, uses the most general type, i.e., an observable on a non-classical quantum space with values in a non-classical quantum space; this is a homomorphism between two non-associative Jordan operator algebras.

A *Jordan algebra*[4] is a linear space $\mathcal{A}$ equipped with a (non-associative) commutative product $\circ$ satisfying $X \circ (Y \circ X^2) = (X \circ Y) \circ X^2$ for all $X, Y \in \mathcal{A}$. A *JB algebra*[4] is a real Jordan algebra $\mathcal{A}$ that is a Banach space with a norm satisfying $\|X \circ Y\| \leq \|X\| \|Y\|$, $\|X^2\| = \|X\|^2$ and $\|X^2\| \leq \|X^2 + Y^2\|$ for all $X, Y \in \mathcal{A}$. The subset $\mathcal{A}_+ := \{X^2 | X \in \mathcal{A}\}$ of a JB algebra $\mathcal{A}$ is a closed convex cone, and a partial ordering is defined via: $X \leq Y \Leftrightarrow Y - X \in \mathcal{A}_+$. For idempotent elements $E$ and $F$, $E \leq F$ is equivalent to $E \circ F = E$. A linear functional $\mu: \mathcal{A} \to \mathbb{R}$ is called *positive* if $\mu(X) \geq 0$ for $X \in \mathcal{A}_+$. A positive linear functional $\mu$ is bounded with $\|\mu\| = \mu(\mathbb{1})$ and is called a *state* if $\mu(\mathbb{1}) = 1$; $\mu$ is *faithful* means that $\mu(X) = 0$ with $X \geq 0$ implies that $X = 0$. A linear map $V$ from a JB algebra $\mathcal{A}$ to another JB algebra is *positive* if $V(X) \geq 0$ for $X \geq 0$. A JB algebra which is the dual of a Banach space is a *JBW algebra*.[4] A JBW algebra has a unit denoted by $\mathbb{1}$ and is generated by its idempotent elements called *events*; the event $E' := \mathbb{1} - E$ is the *negation* of the event $E$. Two events $E$ and $F$ are *orthogonal* if $E \circ F = 0$. The spectral decomposition theorem holds for each element of a JBW algebra.

The *conditional probability* $\mu(F|E)$ of an event $F$ under another event $E$ in a state $\mu$ with $\mu(E) > 0$ has been introduced in Ref. 9, where is has been shown that $\mu(F|E) = \mu(\{E, F, E\})/\mu(E)$. Note that the *triple product* $\{\ ,\ ,\ \}$ in a Jordan algebra is defined as follows: $\{X, Y, Z\} = X \circ (Y \circ Z) - Z \circ (X \circ Y) + Y \circ (Z \circ X)$. For some pairs of events $E$ and $F$, the conditional probability $\mu(F|E)$ does not depend on the underlying state $\mu$ and is then denoted by $\mathbb{P}(F|E)$.

Generally, the equation $\mu(F) = \mu(F|E)\mu(E) + \mu(F|E')\mu(E')$ or the equivalent equation $\mu(\{E, F, E\}) = \mu(E \circ F)$ does not hold; if it does, we write: $E \xrightarrow{\mu} F$. This is a weak state-dependent directed compatibility criterion that has been introduced in Ref. 11 and is always fulfilled in classical probability theory. Several stronger forms of compatibility exist, e.g., the event $E$ and $F$ are called *compatible* if $E \xrightarrow{\mu} F$ and $F \xrightarrow{\mu} E$ both hold for all states $\mu$; in this case, $E$ and $F$ *operator-commute*.[4] With a JBW sub-algebra $\mathcal{A}_1$ of $\mathcal{A}$, we write $\mathcal{A}_1 \xrightarrow{\mu} F$ if $E \xrightarrow{\mu} F$ holds for each event $E$ in $\mathcal{A}_1$, and with two JBW sub-algebras $\mathcal{A}_1$ and $\mathcal{A}_2$ of $\mathcal{A}$, we write $\mathcal{A}_1 \xrightarrow{\mu} \mathcal{A}_2$ if $E \xrightarrow{\mu} F$ holds for each event $E$ in $\mathcal{A}_1$ and each event $F$ in $\mathcal{A}_2$. We shall later see that the condition $\mathcal{A}_1 \xrightarrow{\mu} F$ ensures the existence of a reasonable conditional expectation of $F$ under $\mathcal{A}_1$ in the state $\mu$. This condition is satisfied e.g., when $\mathcal{A}_1$ and $F$ operator-commute, or when $\mathcal{A}$ is the tensor product[11] of $\mathcal{A}_1$ and $\mathcal{A}_2$ with $F \in \mathcal{A}_2$, or when $\mu$ is a trace state.[13]

A linear map (state, homomorphism, observable) on a JBW algebra $\mathcal{A}$ is called *normal* if it is continuous with respect to the weak topology on $\mathcal{A}$ generated by its predual. Normal linear maps are completely additive for any orthogonal family of events. The paper restricts to the study of normal states and normal observables although it would be desirable to include $\sigma$-additive states and standard observables that are $\sigma$-additive and not necessarily normal, but then some methods needed from the theory of JBW algebras would not apply (see also Sec. VII).

The paper is organized as follows: In Sections II and III, some results that are well-known for C*-/W*-algebras are extended to the Jordan operator algebras for later use; these are Kadison's generalized Schwarz inequality and Sakai's Radon-Nikodym theorem. The extension of conditional expectations and Markov processes to the quantum case is introduced in Sections IV and V, respectively. It is shown that a semi-group of positive linear maps is associated with each such Markov process. Finally the generator of this semi-group is studied in Section VI.

Although the motivation for the work presented here primarily stems from a quantum probability model proposed in Refs. 9, 10, and 11, the paper is written in such a way that most of it can be understood without knowledge of that quantum probability model. However, some knowledge of Jordan operator algebras is required, and the monograph Ref. 4 is recommended as an excellent reference.

## II. THE GENERALIZED SCHWARZ INEQUALITY

The following lemma will be needed to prove the generalized Schwarz inequality for positive linear maps between JB algebras.

*Lemma 2.1*: *Let $X_1,...,X_n$ be $n$ positive elements in a JB algebra $\mathcal{A}$ with unit $\mathbb{1}$ such that $\sum_{k=1}^{n} X_k \leq \mathbb{1}$ and $s_1,...,s_n \in \mathbb{R}$. Then*

$$\left(\sum_{k=1}^{n} s_k X_k\right)^2 \leq \sum_{k=1}^{n} s_k^2 X_k .$$

*Proof*: Let $\varphi$ be a state on $\mathcal{A}$. On the direct sum of $n$ copies of $\mathcal{A}$ we consider the positive-semi-definite inner product

$$\langle Y | Z \rangle := \sum_{k=1}^{n} \varphi(\{Y_k, X_k, Z_k\})$$

for $Y = Y_1 \oplus ... \oplus Y_n$ and $Z = Z_1 \oplus ... \oplus Z_n$ with $Y_k, Z_k \in \mathcal{A}$. Then the Cauchy-Schwarz inequality holds [note that $\{Y_k, X_k, Y_k\} \geq 0$ for $Y_k, X_k \in \mathcal{A}$ with $X_k \geq 0$] and, selecting $Y_k := s_k \mathbb{1}$ for $k=1,...,n$ and

$$Z_1 := ... := Z_n := \sum_{k=1}^{n} s_k X_k ,$$

we get:

$$\left(\varphi\left(\left(\sum_{k=1}^{n} s_k X_k\right)^2\right)\right)^2 = \langle Y | Z \rangle^2 \leq \langle Y | Y \rangle \langle Z | Z \rangle$$

$$= \varphi\left(\sum_{k=1}^{n} s_k^2 X_k\right) \varphi\left(\left\{Z_1, \sum_{k=1}^{n} X_k, Z_1\right\}\right)$$

$$\leq \varphi\left(\sum_{k=1}^{n} s_k^2 X_k\right) \varphi(Z_1^2)$$

$$= \varphi\left(\sum_{k=1}^{n} s_k^2 X_k\right) \varphi\left(\left(\sum_{k=1}^{n} s_k X_k\right)^2\right)$$

Hence

$$\varphi\left(\left(\sum_{K=1}^{n} s_k X_k\right)^2\right) \leq \varphi\left(\sum_{K=1}^{n} s_k^2 X_k\right)$$

and, since this holds for every state $\varphi$, the lemma is proved.

The following proposition now provides the extension of Kadison's generalized Schwarz inequality for positive maps between C*-algebras[5] to the more general case of positive maps between JB algebras.

**Proposition 2.2:** *Let V be a positive linear map from a unital JB algebra $\mathcal{M}$ to a unital JB algebra $\mathcal{N}$ with $V(\mathbb{I}) = \mathbb{I}$. Then $(V(X))^2 \leq V(X^2)$ for every $X \in \mathcal{M}$.*

*Proof*: First, we assume that $\mathcal{M}$ is a JBW algebra[4] and that $X$ has the shape $X = \sum_{k=1}^{n} s_k E_k$ with pairwise orthogonal idempotent elements $E_1,...,E_n \in \mathcal{A}$ and $s_1,...,s_n \in \mathbb{R}$. Then $0 \leq V(E_k)$ for $k=1,...,n$ and $\Sigma V(E_k) \leq \mathbb{I}$ such that we can apply Lemma 2.1 getting:

$$(V(X))^2 = \left(\sum_{k=1}^{n} s_k V(E_k)\right)^2 \leq \sum_{k=1}^{n} s_k^2 V(E_k) = V\left(\sum_{k=1}^{n} s_k^2 E_k\right) = V(X^2).$$

Since, due to the spectral theorem, every $X \in \mathcal{M}$ can be approximated in the norm topology by a sequence of elements having the above special shape and since a positive linear map is automatically norm-continuous, $(V(X))^2 \leq V(X^2)$ holds for every $X \in \mathcal{M}$.

We now assume that $\mathcal{M}$ is a JB algebra. Then the double dual spaces $\mathcal{M}^{**}$ and $\mathcal{N}^{**}$ are JBW algebras containing $\mathcal{M}$ and $\mathcal{N}$, respectively, as subalgebras[4], such that Proposition 2.2 follows by applying the above to the map $V^{**}: \mathcal{M}^{**} \to \mathcal{N}^{**}$.

**Corollary 2.3:** *Let V be a bijective positive linear map from a unital JB algebra $\mathcal{M}$ to a unital JB algebra $\mathcal{N}$ with $V(\mathbb{I}) = \mathbb{I}$ such that $V^{-1}$ is positive as well. Then V is a muliplicative homomorphism.*

*Proof*: $(V(X))^2 \leq V(X^2)$ for every $X \in \mathcal{M}$ and $(V^{-1}(Y))^2 \leq V^{-1}(Y^2)$ for every $Y \in \mathcal{N}$. Hence with $Y = V(X)$: $X^2 \leq V^{-1}((V(X))^2)$. Then, since $V$ is positive: $V(X^2) \leq (V(X))^2$. Therefore: $V(X^2) = (V(X))^2$. The identity $2A \circ B = (A+B)^2 - A^2 - B^2$ for $A, B \in \mathcal{M}$ finally implies that $V$ is multiplicative.

## III. THE RADON-NIKODYM THEOREM

We shall now extend one of Sakai's Radon-Nikodym theorems[14] from W*-algebras to JBW algebras. Since all methods needed for the proof of the W*-case are available in the JBW case as well, the proofs are quite similar in these two cases.

**Theorem 3.1:** *Let $\mathcal{A}$ be a JBW algebra and let $\nu$, $\mu$ be two positive normal linear functionals on $\mathcal{A}$ with $\nu \leq \mu$. Then, there is an element $Y \in \mathcal{A}$ with $0 \leq Y \leq \mathbb{I}$ such that $\nu(Z) = \mu(Y \circ Z)$ holds for all $Z \in \mathcal{A}$.*

*Proof*: For $X \in \mathcal{A}$ we define $\mu_X$ via $\mu_X(Z) := \mu(X \circ Z)$ for $Z \in \mathcal{A}$ and consider the set $K := \{\mu_X | X \in \mathcal{A}, 0 \leq X \leq \mathbb{I}\}$ which is a non-empty convex subset of the predual $\mathcal{A}_*$ of $\mathcal{A}$. Moreover, $K$ is compact with respect to the weak topology on $\mathcal{A}_*$ that is generated by $\mathcal{A}$. Note that $\{X \in \mathcal{A}, 0 \leq X \leq \mathbb{I}\}$ is compact and that the multiplication operation with one element fixed is continuous with respect to the weak topology on $\mathcal{A}$ generated by $\mathcal{A}_*$.

All we have to show is that $\nu \in K$. We assume that $\nu \notin K$. From the Hahn-Banach theorem it then follows that there is an element $A \in \mathcal{A}$ and a real number $r$ such that $\nu(A) > r$ and $\mu_X(A) \leq r$ for $X \in \mathcal{A}$ with $0 \leq X \leq \mathbb{I}$. Now let $F$ be the support of the positive part $A_+$ of $A$; these are defined as follows: with $A = \int \lambda \, dE_\lambda$ being the spectral decomposition of $A$, $A_+ := \int_{[0,\infty]} \lambda \, dE_\lambda$ and $F := \int_{[0,\infty]} dE_\lambda$. Then $r \geq \mu_F(A) = \mu(F \circ A) = \mu(A_+) \geq \nu(A_+) \geq \nu(A)$, which contradicts $\nu(A) > r$.

## IV. CONDITIONAL EXPECTATIONS

Let $\mathcal{A}$ be a JBW algebra and $\mathcal{A}_o$ a JBW sub-algebra of $\mathcal{A}$ with $\mathbb{I} \in \mathcal{A}_o$. Let $\mu$ be a normal state on $\mathcal{A}$ and $X \in \mathcal{A}$ with $0 \leq X \leq \mathbb{I}$.

*Definition 4.1*: *An element $Y \in \mathcal{A}_o$ such that $\mu(\{E,X,E\}) = \mu(Y \circ E)$ holds for all events $E$ in $\mathcal{A}_o$ is called a conditional expectation of $X$ under $\mathcal{A}_o$ in the state $\mu$.*

If $F$ is an event in $\mathcal{A}$ and if $Y$ is a conditional expectation of $F$ under $\mathcal{A}_o$ in the state $\mu$, we get that $\mu(F|E)\mu(E) = \mu(Y \circ E)$ holds for all events $E$ in $\mathcal{A}_o$.

*Lemma 4.2*: *If a conditional expectation exists for $X \in \mathcal{A}$ with $0 \leq X \leq \mathbb{I}$, then there is a least one conditional expectation $Y$ with $0 \leq Y \leq \mathbb{I}$.*

*Proof*: Let $Z \in \mathcal{A}_o$ be a conditional expectation with $Z = \int \lambda \, dE_\lambda$ being its spectral decomposition, and define

$$Z_- := -\int_{(-\infty,0)} \lambda \, dE_\lambda, \quad Y := \int_{[0,\infty)} \lambda \, dE_\lambda, \text{ and } E_0 := \int_{(-\infty,0)} dE_\lambda.$$

Then $Z_-, Y, E_0 \in \mathcal{A}_o$. Hence $0 \leq \mu(\{E_0, X, E_0\}) = \mu(Z \circ E_0) = -\mu(Z_-) \leq 0$, i.e. $\mu(Z_-) = 0$. Since $0 \leq Z_-$, we get $\mu(Z_-^2) = 0$, and the Cauchy-Schwarz inequality for states implies that $\mu(Z_- \circ E) = 0$ for all events $E$ in $\mathcal{A}_o$. Therefore $\mu(\{E,X,E\}) = \mu(Z \circ E) = \mu((Z_- + Y) \circ E) = \mu(Y \circ E)$ for all events $E$ in $\mathcal{A}_o$, i.e. $Y$ is a positive conditional expectation of $X$. Repeating now the same procedure for the conditional expectation $\mathbb{I} - Y$ of $\mathbb{I} - X$ finally yields a conditional expectation $Y$ with $0 \leq Y \leq \mathbb{I}$.

**Theorem 4.3:** (i) *A conditional expectation of $X$ under $\mathcal{A}_o$ in the state $\mu$ exists if and only if $\mathcal{A}_0 \xrightarrow{\mu} X$ holds* [*i.e.*, $\mu(\{E,X,E\}) = \mu(E \circ X)$ *for all events* $E \in \mathcal{A}_o$].

(ii) *If $\mathcal{A}_0 \xrightarrow{\mu} X$ holds and if the restriction of $\mu$ to $\mathcal{A}_o$ is faithful, then there is one and only one conditional expectation of $X$ under $\mathcal{A}_o$ in the state $\mu$* [*which shall be denoted by $\mu(X|\mathcal{A}_o)$ in the remaining part of this paper*].

*Proof*: (i) Let $Y$ be a conditional expectation of $X$ under $\mathcal{A}_o$ in the state $\mu$, and let $E$ be an event in $\mathcal{A}_o$. Then $\mu(X) = \mu(Y) = \mu(Y \circ E) + \mu(Y \circ E') = \mu(\{E,X,E\}) + \mu(\{E',X,E'\}) = \mu(X) - 2\mu(X \circ E) + 2\mu(\{E,X,E\})$, where the last equality follows from the identity $\{E',X,E'\} = X - 2X \circ E + \{E,X,E\}$. Therefore $\mu(\{E,X,E\}) = \mu(X \circ E)$.

Now let $\mathcal{A}_0 \xrightarrow{\mu} X$ hold. We then define $\nu(Z) := \mu(X \circ Z)$ for $Z \in \mathcal{A}_0$; $\nu$ is a normal linear functional on $\mathcal{A}_o$. Due to the condition $\mathcal{A}_0 \xrightarrow{\mu} X$, we have $0 \leq \nu(E) \leq \mu(E)$ for all events $E$ in $\mathcal{A}_o$. Therefore $0 \leq \nu \leq \mu$ on $\mathcal{A}_o$. From the Radon-Nikodym theorem we get an element $Y \in \mathcal{A}_o$ with $0 \leq Y \leq \mathbb{I}$ such that $\nu(Z) = \mu(Y \circ Z)$ for all $Z \in \mathcal{A}_0$. Finally, we get that $\mu(Y \circ E) = \mu(X \circ E) = \mu(\{E,X,E\})$ holds for the events $E$ in $\mathcal{A}_o$, where we have again used the assumption $\mathcal{A}_0 \xrightarrow{\mu} X$. Thus, $Y$ is a conditional expectation of $X$ under $\mathcal{A}_o$ in the state $\mu$.

(ii) Now let $\mu$ be faithful on $\mathcal{A}_o$ and $Y_1, Y_2 \in \mathcal{A}_o$ with $0 \leq Y_1, Y_2 \leq \mathbb{I}$ such that $\mu(\{E,X,E\}) = \mu(Y_1 \circ E) = \mu(Y_2 \circ E)$ for all events $E \in \mathcal{A}_o$. Then $\mu((Y_1 - Y_2) \circ E) = 0$ for all events $E \in \mathcal{A}_o$ and, since $\mathcal{A}_o$ is the closed linear hull of its events, we get $\mu((Y_1 - Y_2) \circ Z) = 0$ for all $Z \in \mathcal{A}_o$. Hence $\mu((Y_1 - Y_2)^2) = 0$. The faithfulness now implies that $(Y_1 - Y_2)^2 = 0$. Therefore $Y_1 = Y_2$.

Let the restriction of $\mu$ to $\mathcal{A}_o$ be faithful, and let $\mathcal{A}_0 \xrightarrow{\mu} X, X_1, X_2$ hold with $X, X_1, X_2 \in \mathcal{A}$, $0 \leq X, X_1, X_2 \leq \mathbb{I}$. Lemma 4.2 implies that $0 \leq \mu(X|\mathcal{A}_o) \leq \mathbb{I}$. Obviously we have: $\mu(0|\mathcal{A}_o) = 0$, $\mu(\mathbb{I}|\mathcal{A}_o) = \mathbb{I}$, and $\mu(\alpha X_1 + (1-\alpha) X_2 | \mathcal{A}_o) = \alpha \mu(X_1|\mathcal{A}_o) + (1-\alpha) \mu(X_2|\mathcal{A}_o)$ for $0 \leq \alpha \leq 1$. Moreover, $\mu\big(\mu(X|\mathcal{A}_0)\big|\mathcal{A}_1\big) = \mu\big(X|\mathcal{A}_1\big)$ holds for any other JBW sub-algebra $\mathcal{A}_1 \subseteq \mathcal{A}_o$ with $\mathcal{A}_1 \xrightarrow{\mu} \mu(X|\mathcal{A}_0)$.

The faithfulness of $\mu$ on $\mathcal{A}_o$ is not really a strong restriction; moving over from $\mathcal{A}$ to the JBW algebra $\{D, \mathcal{A}, D\}$ with $D$ being the support of $\mu$, one could even assume that $\mu$ is faithful on $\mathcal{A}$. Note that the support is the smallest event $E$ with $\mu(E) = 1$, which exists for normal states.

We are now in a position to extend the concept of the state-independent conditional probabilities [$\mathbb{P}(F|E)$; see Ref. 9] to the conditional expectations. If $\mu(X|\mathcal{A}_o) = \nu(X|\mathcal{A}_o)$ holds for all normal states $\mu, \nu$ on $\mathcal{A}$, which are faithful on $\mathcal{A}_o$ and satisfy $\mathcal{A}_0 \xrightarrow{\mu} X$ and $\mathcal{A}_0 \xrightarrow{\nu} X$, respectively, and if at least one such state exists, this state-independent conditional expectation is denoted by $\mathbb{E}(X|\mathcal{A}_o)$. The conditional expectations $\mu(F|\mathcal{A}_o)$ and $\mathbb{E}(F|\mathcal{A}_o)$ are elements in $\mathcal{A}_o$ while $\mu(F|E)$ and $\mathbb{P}(F|E)$ are real numbers for events $E$ and $F$ in $\mathcal{A}$. If $E_n$ is a finite or infinite sequence of mutually orthogonal events in $\mathcal{A}$ with $\Sigma E_n = \mathbb{I}$ and $0 < \mu(E_n) < 1$ for each $n$, if $\mathcal{A}_o$ is the sub-algebra generated by the $E_n$, and if $\mathcal{A}_0 \xrightarrow{\mu} F$ holds, we get: $\mu(F|\mathcal{A}_o) = \Sigma \mu(F|E_n) E_n$. If, moreover, the $E_n$ are atoms,[9] $\mathbb{E}(F|\mathcal{A}_o)$ exists and $\mathbb{E}(F|\mathcal{A}_o) = \Sigma \mathbb{P}(F|E_n) E_n$.

With $R: \mathcal{M} \to \mathcal{A}$ being a normal observable, i.e. a JBW-homomorphism from another JBW algebra $\mathcal{M}$ to $\mathcal{A}$, $R(\mathcal{M})$ is a JBW sub-algebra of $\mathcal{A}$. If $R(\mathcal{M}) \xrightarrow{\mu} X$ and if $\mu$ is faithful on $R(\mathcal{M})$, $\mu(X|R(\mathcal{M})) \in R(\mathcal{M})$ is also denoted by $\mu(X|R)$ (or by $\mathbb{E}(X|R)$ in the case of independence of the particular state). If $R$ is injective, there is one and only one element $V_{R,\mu}(X) \in \mathcal{M}$ with $\mu(X|R) = RV_{R,\mu}(X)$. Then $0 \leq V_{R,\mu}(X) \leq \mathbb{I}$.

We now assume that $\mathcal{A}_1$ is a further sub-algebra of $\mathcal{A}$ such that $\mathcal{A}_0 \xrightarrow{\mu} \mathcal{A}_1$ [i.e., $E \xrightarrow{\mu} X$ holds for all events $E$ in $\mathcal{A}_o$ and all $X \in \mathcal{A}_1$] or $R(\mathcal{M}) \xrightarrow{\mu} \mathcal{A}_1$. Then the maps $X \to \mu(X|\mathcal{A}_o)$, $X \to \mu(X|R)$, and $X \to V_{R,\mu}(X)$ [possibly also $X \to \mathbb{E}(X|\mathcal{A}_o)$ and $X \to \mathbb{E}(X|R)$] are convex normal maps from the positive unit ball of $\mathcal{A}_1$ to the positive unit ball of $\mathcal{A}_o$, $R(\mathcal{M})$, and $\mathcal{M}$, respectively. Therefore, each of these maps has a unique extension to a positive normal linear map from $\mathcal{A}_1$ to $\mathcal{A}_o$, $R(\mathcal{M})$, and $\mathcal{M}$, respectively. Thus, $\mu(X|\mathcal{A}_o)$, $\mu(X|R)$, and $V_{R,\mu}(X)$ [possibly also $\mathbb{E}(X|\mathcal{A}_o)$ and $\mathbb{E}(X|R)$] are defined for all $X \in \mathcal{A}_1$.

Differing from the notation used here, a positive linear map $\pi: \mathcal{A} \to \mathcal{A}_o$, $X \to \pi(X)$ with $\mathcal{A}_o \subseteq \mathcal{A}$, $\pi = \pi^2$ and $\pi(\mathbb{I}) = \mathbb{I}$ is sometimes called a conditional expectation [e.g., Refs. 1 and 14]; in Ref. 4, it is shown that then $\pi(X \circ Y) = \pi(X) \circ Y$ for $X \in \mathcal{A}$ and $Y \in \mathcal{A}_o$. If $\mathcal{A}$ is a JBW algebra with a faithful trace state (e.g., a JBW algebra with a finite dimension, or a type $\mathrm{II}_1$ factor), then it follows from Theorem 4.3 that a positive linear map $\pi: \mathcal{A} \to \mathcal{A}$ with $\pi = \pi^2$, $\pi(\mathbb{I}) = \mathbb{I}$ and $\pi(\mathcal{A}) = \mathcal{A}_o$ exists for each JBW sub-algebra $\mathcal{A}_o \subseteq \mathcal{A}$; note that, with $\mu$ being a trace state, $E \xrightarrow{\mu} F$ holds for all events $E$ and $F$ in $\mathcal{A}$, and define $\pi(X) := \mu(X|\mathcal{A}_0)$ for $X \in \mathcal{A}$.

## V. THE MARKOV PROCESS

Let $\mathcal{A}$ be a JBW algebra. Let $R_s: \mathcal{M}_s \to \mathcal{A}$ ($s \in S$) be a family of normal observables with $\mathcal{M}_s$ being further JBW algebras. Typical examples of the index set $S$ are intervals, e.g., $[0,\infty]$ or $[-\infty,\infty]$, the integers or the non-negative integers $\{0,1,2,...\}$. The JBW sub-algebra of $\mathcal{A}$ that is generated by $\bigcup_{s' \leq s} R_{s'}(\mathcal{M}_{s'})$ is denoted by $\mathcal{A}_{\leq s}$ for $s \in S$.

Definition 5.1: *The family of normal observables $R_s$ ($s \in S$) together with a faithful normal state $\mu$ on $\mathcal{A}$ is now called a normal Markov process if*

(i) $\mathcal{A}_{\leq s} \xrightarrow{\mu} R_{s'}(\mathcal{M}_{s'})$ *for $s<s'$,*

and

(ii) $\mu(X|\mathcal{A}_{\leq s}) = \mu(X|R_s)$ *for $s<s'$ and $X \in R_{s'}(\mathcal{M}_{s'})$*

*both hold.*

Condition (i) is a weak directed compatibility criterion for the family of observables $R_s$ ($s \in S$) under the fixed state $\mu$. Within classical probability theory, it is meaningless, since generally holding. Condition (ii) is the Markov property, meaning that a Markov process is memoryless; the future behavior of the process after time $s$ depends only on its behavior at time $s$ and not on the process history before time $s$.

We now assume a normal Markov process with each $R_s$ ($s \in S$) being injective; then $V_{R_s,\mu}(R_{s'}(Y)) \in \mathcal{M}_s$ for $Y \in \mathcal{M}_{s'}$ and $s<s'$ (see Sec. IV). Hence, the composition $V_{s,s'} := V_{R_s,\mu}R_{s'}$ is a positive normal map $\mathcal{M}_{s'} \to \mathcal{M}_s$ with $V_{s,s'}(\mathbb{I}) = \mathbb{I}$. Please, keep in mind that the $V_{s,s'}$ depend on the underlying state $\mu$, although this is not shown in the nomenclature.

**Theorem 5.2:** (i) $\mu^{R_s}(V_{s,s'}(Y)) = \mu^{R_{s'}}(Y)$ *for $Y \in \mathcal{M}_{s'}$ and $s<s'$; i.e., $V_{s,s'}$ transfers the probability distribution of $R_s$ on $\mathcal{M}_s$ into the one of $R_{s'}$ on $\mathcal{M}_{s'}$. Note that $\mu^R$ with $\mu^R(E) := \mu(R(E))$ is the distribution of the observable $R$ in the state $\mu$.*
(ii) $V_{s,s'}V_{s',s''} = V_{s,s''}$ *for $s,s',s'' \in S$ with $s < s' < s''$.*

*Proof*: (i) $\mu^{R_s}(V_{s,s'}(Y)) = \mu((R_s V_{R_s,\mu} R_{s'})(Y)) = \mu(\mu(R_{s'}(Y)|R_s)) = \mu(R_{s'}(Y)) = \mu^{R_{s'}}(Y)$.
(ii) Let $X \in R_{s''}(\mathcal{M}_{s''})$. Note that $\mathcal{A}_{\leq s} \xrightarrow{\mu} \mu(X|R_{s'}) = \mu(X|A_{\leq s'})$. Therefore

$$\mu(X|R_s) = \mu(X|\mathcal{A}_{\leq s}) = \mu(\mu(X|\mathcal{A}_{\leq s'})|\mathcal{A}_{\leq s}) = \mu(\mu(X|R_{s'})|\mathcal{A}_{\leq s}) = \mu(\mu(X|R_{s'})|R_s),$$

where the Markov property has been applied several times. With $X = R_{s''}(Y)$ we get: $(R_s V_{R_s,\mu} R_{s''})(Y) = (R_s V_{R_s,\mu} R_{s'} V_{R_{s'},\mu} R_{s''})(Y)$ for all $Y \in M_{s''}$ and, due to the injectivity of $R_s$: $V_{R_s,\mu} R_{s''} = V_{R_s,\mu} R_{s'} V_{R_{s'},\mu} R_{s''}$. The left-hand side of this last equation is identical with $V_{s,s''}$ and the right-hand side is identical with $V_{s,s'}V_{s',s''}$.

Part (ii) of the theorem is the quantum version of the Chapman-Kolmogorov equation in classical probability theory. It holds if the Markov property is satisfied. Part (i) is valid more generally [i.e., if $R_s(\mathcal{M}_s) \xrightarrow{\mu} R_{s'}(\mathcal{M}_{s'})$ is satisfied for $s<s'$]. The approach to Markov processes, presented here, is very similar to the one in classical mathematical probability theory (e.g., Ref. 8). The adaptation to the quantum case becomes possible due to the compatibility criterion (i) in Definition 5.1. The faithfulness of the underlying state $\mu$ and the injectivity of the $R_s$ are technical assumptions to avoid the difficulties involved with the $\mu$-almost-everywhere equivalence classes that are used in mathematical probability theory. Some further discussion of the assumptions will follow in the concluding remarks.

We call a Markov process *reversible*, if each $V_{s,s'}$ has an inverse $V_{s,s'}^{-1}$ and if this inverse is a positive map. Then $V_{s,s'}^{-1}$ transfers the probability distribution of $R_{s'}$ on $\mathcal{M}_{s'}$ into the one of $R_s$ on $\mathcal{M}_s$. Without $V_{s,s'}^{-1}$ being positive, it would not transfer states to states, but to linear functionals that are not necessarily positive. The generalized Schwarz inequality (Corollary 2.3) implies that the $V_{s,s'}$ are multiplicative isomorphisms in the case of a reversible Markov process.

## VI. THE DYNAMICAL GROUP AND ITS GENERATOR

We now assume a Markov process such that each $R_s$ ($s \in S$) is injective and $\mathcal{M}_s = \mathcal{M}$ for all $s$. Let $Aut(\mathcal{M})$ denote the automorphism group of $\mathcal{M}$ and let $Pos(\mathcal{M})$ be the set of all positive normal linear maps from $\mathcal{M}$ to $\mathcal{M}$ that map the unit element to itself; $Pos(\mathcal{M})$ is a semi-group, but an inverse need not exist for an element of $Pos(\mathcal{M})$. The $V_{s,s'}$ now lie in $Pos(\mathcal{M})$, and if the Markov process is reversible, they lie in $Aut(\mathcal{M})$.

With the index set $S$ being one of the sets $(-\infty,\infty)$, $[0,\infty)$, $\{0,1,2,...\}$ or $\{...,-1,0,1,2,...\}$ and with $\mathcal{M}_s = \mathcal{M}$ for all $s \in S$, we call a Markov process *stationary*, if $\mu(R_{s'}(Y)|R_s) = \mu(R_{t'}(Y)|R_t)$ for $s'-s=t'-t$ [$s,t \in S$, $s<s'$, $t<t'$ and $Y \in \mathcal{M}$]. Then $V_{s,s'} = V_{t,t'}$ for $s'-s=t'-t$, and we can define: $V_{t,t'} =: V_{t'-t}$. For these $V_t$ we have: $V_s V_t = V_{s+t}$. Note that $V_{s,s}$ ($s = s'$) as well as $V_0$ ($t=0$) have not been defined so far, and we now define $V_{s,s}$ and $V_0$ to be the identity map on $\mathcal{M}$. Thus, with $S=[0,\infty)$ and a stationary Markov process, the $V_t$ form a dynamical semi-group in $Pos(\mathcal{M})$; with $S=(-\infty,\infty)$ and a stationary and reversible Markov process, the $V_t$ form a dynamical group in $Aut(\mathcal{M})$.

We shall now briefly consider the generators of such groups, but will not go into the technical details of the different kinds of convergence, since most of it is well-known - if not for JBW algebras, then at least for the C*-/W*-algebras. If

$$L := \frac{d}{dt} V_t \bigg|_{t=0}$$

exists (convergence in some sense assumed) for a stationary Markov process, $L$ is a linear operator $\mathcal{M} \to \mathcal{M}$ (or possibly defined only on a dense subset of $\mathcal{M}$) and the following differential equation holds:

$$\frac{d}{dt} V_t = LV_t .$$

Then $V_t = \exp(tL)$. $L$ is called the *generator* of the dynamical (semi-)group $V_t$. From the generalized Schwarz inequality we get for $Y \in \mathcal{M}$:

$$V_t(Y^2) - Y^2 \geq (V_t(Y))^2 - Y^2 = (V_t(Y) + Y) \circ (V_t(Y) - Y)$$

and therefore (note that $V_o$ is the identity):

$$L(Y^2) = \frac{d}{dt}V_t(Y^2)\bigg|_{t=0} = \lim_{t \downarrow 0}\frac{1}{t}(V_t(Y^2) - Y^2) \geq \lim_{t \downarrow 0}(V_t(Y) + Y) \circ \lim_{t \downarrow 0}\frac{1}{t}(V_t(Y) - Y) = 2Y \circ L(Y).$$

In the reversible case, each "≥" can be replaced by "=" (Corollary 2.3). Then $L(Y^2) = 2Y \circ L(Y)$ for $Y \in \mathcal{M}$. Linear maps $L$ satisfying this equation are called *derivations*. We call a linear map $L$ satisfying $L(Y^2) \geq 2Y \circ L(Y)$ for $Y \in \mathcal{M}$ a *dissipation*.

Thus, we have shown that the generator of the dynamical group associated with a stationary Markov process is a dissipation, and is a derivation if the Markov process is stationary and reversible. With $\mathcal{M}$ being the self-adjoint part of a W*-algebra and with the inner derivation $L(X) := i[H,X]$ for $X \in \mathcal{M}$ [with some $H \in \mathcal{M}$], we get

$$\frac{d}{dt}V_t(X) = i[H,V_t(X)] \text{ and } V_t(X) = e^{itH}Xe^{-itH}.$$

This provides the Schroedinger equation and its solution as a very special case of a more general approach.

Derivations on C*-algebras are studied in Refs. 1 and 2. The above definition of a dissipation differs from the dissipations studied in Refs. 2 and 3. In Ref. 2, a linear map δ on a *-subalgebra $\mathcal{A}_o$ of a C*-algebra, satisfying δ(X*X) ≤ δ(X*)X + X*δ(X) for $X \in \mathcal{A}_o$, is called a dissipation. This implies, but is not equivalent to δ(X²) ≤ δ(X)X + Xδ(X) = 2X ∘ δ(X) for all self-adjoint X in $\mathcal{A}_o$. Other authors use the reversed inequality δ(X*X) ≥ δ(X*)X + X*δ(X) for the definition of a dissipation. Note that our definition is based on this reversed inequality which, moreover, is required to hold for the self-adjoint elements only (which form the JB algebra).

Note that other authors immediately define a quantum Markov process as a pair consisting of a W*-algebra and a dynamical semi-group of (completely) positive normal maps on this W*-algebra, without starting from a stochastic process given as a family of observables and implicitly assuming the stationarity.

## VII. CONCLUDING REMARKS

The concept of the classical conditional expectations has been extended to the quantum case, using the framework of Jordan operator algebras. An important condition for the existence of the conditional expectations is given by a certain weak compatibility criterion that was introduced earlier in Ref. 11. With these concepts, it has been possible to partly extend the classical equivalence between two different ways of describing a Markov process to the quantum case. Starting from a Markov process given as a family of observables, we have derived the positive maps $V_{s,s'}$. In classical probability theory, the reverse is also possible; a Markov process consisting of a family of random variables can be reconstructed from a system of Markovian kernels by using the concept of product σ-algebras. The problem of finding a satisfying analogue of these product σ-algebras for the quantum case has been addressed, but only partly been solved in Ref. 11.

The appropriate framework for this approach to quantum Markov processes are Jordan operator algebras, but neither the only norm-complete JB algebras nor the weakly complete JBW algebras are really satisfying. The JB algebras do not contain sufficiently many idempotent elements (quantum events). If the theory is based upon JBW algebras (as here in this paper), an important example, the algebra consisting of the measurable real-valued functions on a measurable space (the real-valued random variables of classical probability theory) is ruled out. This algebra can be embedded in a JBW algebra [even in a W*-algebra[7]], however, one is rather reluctant to work with this abstractly constructed JBW algebra instead of the well-understood algebra of measurable real functions. What is needed is a theory of monotone-sequentially complete JB algebras, similar to the one of the monotone-sequentially complete C*-algebras studied by Kadison,[6] Kehlet[7] and Pedersen.[12] This theory must include a Gleason-type theorem and a Radon-Nikodym theorem for σ-additive states. The Gleason-type theorem is needed for proving that unique conditional probabilities exist for σ-additive states defined on the system of events [as done for the JBW case in Ref. 9], and the Radon-Nikodym theorem is required for showing that the conditional expectations exist for the σ-additive states.